\documentclass{article}
\usepackage[dvipsnames,svgnames,x11names]{xcolor}
\usepackage{amsmath,amssymb}
\usepackage{lmodern}
\usepackage[T1]{fontenc}
\usepackage[utf8]{inputenc}
\usepackage{textcomp}

\IfFileExists{upquote.sty}{\usepackage{upquote}}{}
\IfFileExists{microtype.sty}{
  \usepackage[]{microtype}
  \UseMicrotypeSet[protrusion]{basicmath} 
}{}
\makeatletter
\@ifundefined{KOMAClassName}{
  \IfFileExists{parskip.sty}{%
    \usepackage{parskip}
  }{
    \setlength{\parindent}{0pt}
    \setlength{\parskip}{6pt plus 2pt minus 1pt}}
}{
  \KOMAoptions{parskip=half}}
\makeatother
\usepackage{xcolor}
\usepackage{graphicx}
\makeatletter
\def\maxwidth{\ifdim\Gin@nat@width>\linewidth\linewidth\else\Gin@nat@width\fi}
\def\maxheight{\ifdim\Gin@nat@height>\textheight\textheight\else\Gin@nat@height\fi}
\makeatother
\setkeys{Gin}{width=\maxwidth,height=\maxheight,keepaspectratio}
\makeatletter
\def\fps@figure{htbp}
\makeatother
\providecommand{\tightlist}{%
  \setlength{\itemsep}{0pt}\setlength{\parskip}{0pt}}
\NewDocumentCommand\citeproctext{}{}
\NewDocumentCommand\citeproc{mm}{%
  \begingroup\def\citeproctext{#2}\cite{#1}\endgroup}
\makeatletter
 \let\@cite@ofmt\@firstofone
 \def\@biblabel#1{}
 \def\@cite#1#2{{#1\if@tempswa , #2\fi}}
\makeatother
\newlength{\cslhangindent}
\newlength{\csllabelwidth}
\newenvironment{CSLReferences}[2] 
 {\begin{list}{}{%
  \setlength{\itemindent}{0pt}
  \setlength{\leftmargin}{0pt}
  \setlength{\parsep}{0pt}
  \ifodd #1
   \setlength{\leftmargin}{\cslhangindent}
   \setlength{\itemindent}{-1\cslhangindent}
  \fi
  \setlength{\itemsep}{#2\baselineskip}}}
 {\end{list}}
\usepackage{calc}

\IfFileExists{bookmark.sty}{\usepackage{bookmark}}{\usepackage{hyperref}}
\IfFileExists{xurl.sty}{\usepackage{xurl}}{} 
\hypersetup{
  pdftitle={phylo2vec: a library for vector-based phylogenetic tree
manipulation},
  pdfauthor={Neil Scheidwasser, Ayush Nag, Matthew J Penn, Anthony MV
Jakob, Frederik Mølkjær Andersen, Mark P Khurana, Landung Setiawan,
David A Duchêne, Samir Bhatt},
  colorlinks=true,
  linkcolor={Maroon},
  filecolor={Maroon},
  citecolor={Blue},
  urlcolor={Blue},
  pdfcreator={LaTeX via pandoc}}

\title{phylo2vec: a library for vector-based phylogenetic tree
manipulation}

\definecolor{c53baa1}{RGB}{83,186,161}
\definecolor{c202826}{RGB}{32,40,38}

\usepackage[affil-it]{authblk}
\usepackage{orcidlink}
\author[1%
  *%
  ]{Neil Scheidwasser%
    \,\orcidlink{0000-0001-9922-0289}\,%
    }
\author[2%
  *%
  ]{Ayush Nag%
    }
\author[1%
  ]{Matthew J Penn%
    \,\orcidlink{0000-0001-8682-5393}\,%
    }
\author[4%
  ]{Anthony MV Jakob%
    \,\orcidlink{0000-0002-0996-1356}\,%
    }
\author[1%
  ]{Frederik Mølkjær Andersen%
    \,\orcidlink{0009-0004-4071-3707}\,%
    }
\author[1%
  ]{Mark P Khurana%
    \,\orcidlink{0000-0002-1123-7674}\,%
    }
\author[2%
  ]{Landung Setiawan%
    \,\orcidlink{0000-0002-1624-2667}\,%
    }
\author[1%
  ]{David A Duchêne%
    \,\orcidlink{0000-0002-5479-1974}\,%
    }
\author[1,3%
  \ensuremath\mathparagraph]{Samir Bhatt%
    \,\orcidlink{0000-0002-0891-4611}\,%
    }

\affil[1]{Section of Health Data Science and AI, University of
Copenhagen, Copenhagen, Denmark%
  }
\affil[2]{eScience Institute, University of Washington, Seattle, United
States%
  }
\affil[3]{MRC Centre for Global Infectious Disease Analysis, Imperial
College London, London, United Kingdom%
  }
\affil[4]{Independent researcher%
  }
\affil[$\mathparagraph$]{Corresponding author: %
}
\affil[*]{These authors contributed equally.}
\date{14 May 2025}

\begin{document}
\maketitle

\section{Summary}\label{summary}

Phylogenetics is a fundamental component of evolutionary analysis
frameworks in biology (\citeproc{ref-yang2014}{Yang, 2014}) and
linguistics (\citeproc{ref-atkinson2005}{Atkinson \& Gray, 2005}).
Recently, the advent of large-scale genomics and the SARS-CoV-2 pandemic
has highlighted the necessity for phylogenetic software to handle large
datasets (\citeproc{ref-attwood2022}{Attwood et al., 2022};
\citeproc{ref-kapli2020}{Kapli et al., 2020};
\citeproc{ref-khurana2024}{Khurana et al., 2024};
\citeproc{ref-kraemer2025}{Kraemer et al., 2025}). While significant
efforts have focused on scaling optimisation algorithms
(\citeproc{ref-demaio2023}{De Maio et al., 2023};
\citeproc{ref-sanderson2021}{Sanderson, 2021};
\citeproc{ref-turakhia2021}{Turakhia et al., 2021}), visualization
(\citeproc{ref-sanderson2022}{Sanderson, 2022}), and lineage
identification (\citeproc{ref-mcbroome2024}{McBroome et al., 2024}), an
emerging body of research has been dedicated to efficient
representations of data for genomes
(\citeproc{ref-deorowicz2023}{Deorowicz et al., 2023}) and phylogenetic
trees (\citeproc{ref-chauve2025}{Chauve et al., 2025};
\citeproc{ref-penn2024}{Penn et al., 2024};
\citeproc{ref-richman2025}{Richman et al., 2025}). Compared to the
traditional Newick format which represents trees using strings of nested
parentheses (\citeproc{ref-felsenstein2004}{Felsenstein, 2004}), modern
tree representations utilize integer vectors to define the tree topology
traversal. This approach offers several advantages, including easier
manipulation, increased memory efficiency, and applicability to machine
learning.

\begin{figure}
\centering
\includegraphics{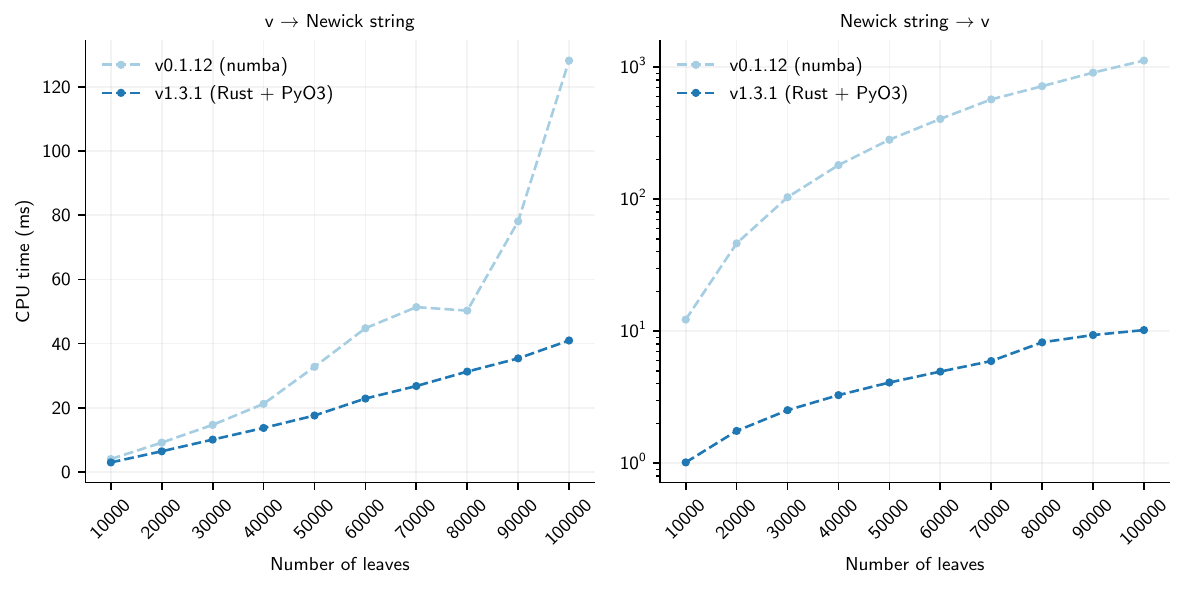}
\caption{Benchmark times for converting a phylo2vec vector to a Newick
string (left) and vice versa (right). Execution time was measured over
at least 20 runs per size, comparing Python functions in the latest
release (via Rust bindings with
\href{https://github.com/PyO3/pyo3}{PyO3}) against the previous release
(\citeproc{ref-penn2024}{Penn et al., 2024}) based on \texttt{Numba}
(\citeproc{ref-lam2015}{Lam et al., 2015}). All benchmarks were
performed on a workstation equipped with an AMD Ryzen Threadripper PRO
5995WX (64 cores, 2.7 GHz) and 256 GB of RAM. \label{fig:benchmarks}}
\end{figure}

Here, we present the latest release of \texttt{phylo2vec} (or
Phylo2Vec), a high-performance software package for encoding,
manipulating, and analysing binary phylogenetic trees. At its core, the
package is based on the phylo2vec (\citeproc{ref-penn2024}{Penn et al.,
2024}) representation of binary trees, and is designed to enable fast
sampling and tree comparison. This release features a core
implementation in Rust for improved performance and memory efficiency
(\autoref{fig:benchmarks}), with wrappers in R and Python (superseding
the original release (\citeproc{ref-penn2024}{Penn et al., 2024})),
making it accessible to a broad audience in the bioinformatics
community.

\section{Statement of need}\label{statement-of-need}

The purpose of the \texttt{phylo2vec} library is threefold. First, it
provides robust phylogenetic tree manipulation in Rust, complementing
other efforts such as \texttt{light\_phylogeny}
(\citeproc{ref-duchemin2018}{Duchemin et al., 2018}) for reconciled
phylogenies (\citeproc{ref-nakhleh2013}{Nakhleh, 2013}), and
\texttt{rust-bio} (\citeproc{ref-koster2016}{Köster, 2016}), which does
not yet cover phylogenetics. Second, it complements existing libraries
such as \texttt{ape} (\citeproc{ref-paradis2019}{Paradis \& Schliep,
2019}) in R, and \texttt{ete3} (\citeproc{ref-huerta2016}{Huerta-Cepas
et al., 2016}) and \texttt{DendroPy} (\citeproc{ref-Moreno2024}{Moreno
et al., 2024}) in Python, by providing fast tree sampling, fast tree
comparison and efficient tree data compression
(\citeproc{ref-penn2024}{Penn et al., 2024}). Third, the phylo2vec
representation offers a pathway to using new optimisation frameworks for
phylogenetic inference. A notable example is GradME
(\citeproc{ref-penn2023}{Penn et al., 2023}), a gradient descent-based
algorithm which uses a continuous relaxation of the phylo2vec
representation.

\section{Features}\label{features}

The presented release of \texttt{phylo2vec} addresses optimisations
limitations of (\citeproc{ref-penn2024}{Penn et al., 2024}) with
\(\mathcal{O}(n \log n)\) implementations for vector-to-Newick and
Newick-to-vector conversions, leveraging Adelson-Velsky and Landis (AVL)
trees (\citeproc{ref-adelson1962}{Adelson-Velsky \& Landis, 1962}) and
Fenwick trees (\citeproc{ref-fenwick1994}{Fenwick, 1994}), respectively.

New features include an extension of the vector representation to
support branch length annotations, leaf-level operations (pruning,
placement, MRCA identification), fast cophenetic distance matrix
calculation, and various optimisation schemes based on phylo2vec tree
representations, notably hill-climbing (\citeproc{ref-penn2024}{Penn et
al., 2024}) and GradME (\citeproc{ref-penn2023}{Penn et al., 2023}). We
also propose a likelihood function for Bayesian MCMC inference that
leverages tree representation similarities with \texttt{BEAGLE}
(\citeproc{ref-ayres2012}{Ayres et al., 2012};
\citeproc{ref-suchard2009}{Suchard \& Rambaut, 2009}). Finally,
user-friendliness is enhanced with step-by-step demos of phylo2vec's
representations and core functions.

\section{Maintenance}\label{maintenance}

A strong focus of this release is to support long-term maintenance
through implementing recommended software practices into its project
structure and development workflow. The project is structured with a
Rust API containing core algorithms with language bindings to avoid
tight coupling and enable easy language additions. Additionally, we have
established a robust continuous integration (CI) pipeline using GitHub
Actions, which features:

\begin{itemize}
\tightlist
\item
  Unit test frameworks for Rust (\href{https://crates.io}{cargo}),
  Python (\href{https://github.com/pytest-dev/pytest}{pytest}), and R
  (testthat (\citeproc{ref-wickham2011}{Wickham, 2011}))
\item
  Benchmarking on the Rust code
  (\href{https://github.com/bheisler/criterion.rs}{criterion}) and its
  Python bindings
  (\href{https://github.com/ionelmc/pytest-benchmark}{pytest-benchmark})
\end{itemize}

Lastly, to complement Jupyter Notebook demos, comprehensive
documentation is provided using \href{https://jupyterbook.org}{Jupyter
Book} and \href{https://about.readthedocs.com/}{Read The Docs}.

\section{Acknowledgements}\label{acknowledgements}

S.B. acknowledges funding from the MRC Centre for Global Infectious
Disease Analysis (reference MR/X020258/1), funded by the UK Medical
Research Council (MRC). This UK funded award is carried out in the frame
of the Global Health EDCTP3 Joint Undertaking. S.B. acknowledges support
from the Danish National Research Foundation via a chair grant (DNRF160,
also supporting N.S. and M.P.K.), The Eric and Wendy Schmidt Fund For
Strategic Innovation via the Schmidt Polymath Award (G-22-63345, also
supporting M.J.P. and F.M.A.), and the Novo Nordisk Foundation via The
Novo Nordisk Young Investigator Award (NNF20OC0059309). D.A.D. is
supported by a Data Science - Emerging researcher award from Novo
Nordisk Fonden (NNF23OC0084647). The authors also thank Madeline Gordon
for her help in implementing functions related to the matrix format.

\section*{References}\label{references}
\addcontentsline{toc}{section}{References}

\phantomsection\label{refs}

\end{document}